\documentclass[a4paper,12pt]{article}
\pdfoutput=1
\usepackage{cite}
\usepackage{cutwin}
\usepackage[utf8]{inputenc}
\usepackage{amsmath,amsfonts,amscd,amssymb,mathrsfs}
\DeclareMathOperator{\Tr}{Tr}

\usepackage[T1]{fontenc}
\usepackage[english]{babel}
\usepackage{epsfig}
\usepackage{changebar}
\usepackage{lmodern} 
\usepackage[all]{xy}
\usepackage{mathtools}
\usepackage{MnSymbol}
\usepackage{fancybox}
\usepackage{wrapfig}
\usepackage{cite}
\usepackage{hyperref}
\usepackage{fancyhdr}
\usepackage{vmargin} 
\usepackage[nottoc, notlof, notlot]{tocbibind}
\usepackage{graphicx}
\usepackage{caption}
\usepackage{subcaption}
\usepackage{titlesec}
\usepackage{color}
\usepackage{stmaryrd}
\usepackage[most]{tcolorbox}
\usepackage{wrapfig,lipsum,booktabs}
\usepackage{listings}
\usepackage{MnSymbol}

\definecolor{dkgreen}{rgb}{0,0.6,0}
\definecolor{gray}{rgb}{0.5,0.5,0.5}
\definecolor{mauve}{rgb}{0.58,0,0.82}

\titleformat{\chapter}[display]   
{\normalfont\huge\bfseries}{\chaptertitlename\ \thechapter}{20pt}{\Huge}   
\titlespacing*{\chapter}{0pt}{-50pt}{40pt}

\titlespacing*{\section}{0pt}{0.2\baselineskip}{\baselineskip}

\DeclareMathOperator*{\tr}{Tr}
\setmarginsrb{2cm}{2cm}{2cm}{2cm}{14mm}{6mm}{0mm}{10mm}
\newcommand{\imineq}[2]{\vcenter{\hbox{\includegraphics[height=#2ex]{#1}}}}

\DeclareMathOperator*{\T}{\boldsymbol{T}}
\DeclareMathOperator*{\Beta}{\boldsymbol{\beta}}
\DeclareMathOperator*{\Q}{\boldsymbol{Q}}
\DeclareMathOperator*{\sgn}{sign}

\begin{document}
\begin{center}
{\Large {\bf Cumulants of conserved charges in GGE and cumulants of total transport in GHD: exact summation of matrix elements?}}

\vspace{1cm}

{\large Dinh-Long Vu}
\vspace{0.2cm}

{\small\em
 Institut de Physique Th\'eorique, CEA Saclay, Gif-Sur-Yvette, 91191, France}
\end{center}

\vspace{1cm}

\noindent We obtain the cumulants of conserved charges in Generalized Gibbs Ensemble (GGE) by a direct summation of their finite-particle matrix elements. The Gaudin determinant that describes the norm of Bethe states is written as a sum over forests by virtue of the matrix-tree theorem. The aforementioned cumulants are then given by a sum over tree-diagrams whose Feynman rules involve simple Thermodynamic Bethe Ansatz (TBA) quantities.  The internal vertices of these diagrams have the interpretation of virtual particles that carry anomalous corrections to bare charges. Our derivation follows closely the spirit of recent works \cite{Kostov:2018dmi, Kostov:2018ckg}. We also conjecture that the cumulants of total transport in Generalized Hydrodynamics (GHD) are given by the same diagrams up to minor modifications. These cumulants play a central role in large deviation theory  and were obtained in \cite{2018arXiv181202082M} using  linear fluctuating hydrodynamics at Euler scale. We match  our conjecture with the result of \cite{2018arXiv181202082M} up to the fourth cumulant. This highly non-trivial matching provides a strong support for our conjecture.
\vspace{1cm}

{\ }\hfill
%\today
\newpage
\section*{Introduction}
%steady state if exists: described by GGE, non-equilibrium: breaking time reversal symmetry

The complex out-of-equilibrium physics \cite{Eisert:2014jea}  of many-body quantum systems calls for suitable testing grounds. One dimensional integrable models have proven to be promising candidates in view of their unorthodox relaxation \cite{PhysRevLett.98.050405,PhysRevLett.106.140405}, possibility for analytical computation \cite{Lieb:1963rt,Yang:1968rm} and realization in cold atom experiments \cite{PhysRevLett.100.090402,PhysRevA.88.031603,Langen207}. To study their transport properties, a  theoretical framework called Generalized Hydrodynamics (GHD) has recently been  developed \cite{Castro-Alvaredo:2016cdj,PhysRevLett.117.207201}. GHD aims at providing a hydrodynamic description of integrable systems out of equilibrium while taking into account their infinite number of integrals of motion. It suggests that at long time the system can be regarded as a collection of mesoscopic-sized fluid cells. The state at each fluid cell is subjected to  local entropy maximization  and is described by a local Generalized Gibbs Ensemble (GGE).  One can equivalently characterize a state by its set of conserved charge averages. The current carried by the state can be deduced from the microscopic continuity equation and the local equilibrium assumption. Quantities involving conserved charges and currents play a central role in the dynamics at Euler scale \cite{SciPostPhys.3.6.039,Doyon:2017vfc}.

A series of recent papers \cite{Kostov:2018ckg,Kostov:2018dmi,10.21468/SciPostPhys.6.2.023} aims at deriving ensemble average of observables from an exact summation of  their matrix elements. For the partition function (open and closed systems) and one point function of local operators, it was surprisingly found that some thermodynamic structures already manifest  themselves in finite particle matrix elements. In this paper we extend  this idea to the cumulants of conserved charges in a GGE. We also conjecture that the cumulants of  total transport currents in a  stationary state possess the same structure. Hereupon we mean by a total current the time integrated current scaled by inverse time, taken in the infinite time limit. We expect that our derivation can provide insights on other quantities like the Drude weight or dynamical correlation functions at Euler scale.

Before presenting the novel aspects of this paper let us first recall the known methods to obtain these cumulants. The Thermodynamic Bethe Ansatz (TBA) of GGE has been established in \cite{Mossel:2012vp} where the authors also obtained the average of the conserved charges and their covariance. Higher cumulants are given by higher derivatives of the TBA free energy. Although this procedure is straight forward, the obtained expressions cannot be considered as explicit. The reason is that they are written in terms of higher derivatives of the TBA free pseudo-energy, each one of which is the solution of an integral equation that involves all the lower derivatives. On the other hand, cumulants of total currents in stationary states have only been recently considered. The covariance matrix is called the Drude self-weight in \cite{Doyon_2017} for its similar appearance with the conventional Drude weight. In the same paper the authors also computed this quantity by combining the current sum rule  \cite{Mendl_2015} with the long wavelength limit of the charge-charge correlation function \cite{SciPostPhys.1.2.015}. The cumulant generating function of the total currents (also known as the full counting statistics) plays an important role in large deviation theory and has been studied in \cite{2018arXiv181202082M}. By using linear fluctuating hydrodynamics at Euler scale, the authors found a functional equation satisfied by this function. This constraint is tight enough for  individual cumulants to be extracted, although case-dependent manipulations are required for their explicit expressions.

Let us now summarize the results of this paper. In the first part we show that the cumulants of  conserved charges in a GGE can be written as a sum over tree diagrams. The $n^\text{th}$ cumulant is given by a sum over rooted trees with $n$ leaves, each of which carries a conserved charge. The internal vertices play the role of virtual particles that carry anomalous corrections to the bare charges. The weight of vertices and propagators are conveniently expressed in terms of TBA quantities. This derivation employs the same technique of \cite{Kostov:2018ckg,Kostov:2018dmi}, namely expanding the Gaudin norm of Bethe states into a sum over forests. In the second part, we conjecture that the cumulants of total currents are given by the exact same trees, up to only minor modifications of the weights. We compare our conjecture with the result of \cite{2018arXiv181202082M} and find perfect agreement up to the fourth cumulant.

Our systematic treatment not only reduces the computational complexity but also improves the  conceptual understanding of these cumulants. First, the simple combinatorial structure of the cumulants of total currents potentially translates into an analytic property of the full counting statistics. It is interesting  to find a new relation in addition to the one established in  \cite{2018arXiv181202082M}. Second, such structure provides hints about what the corresponding matrix elements would look like. For the current average, this line of idea has been exploited in recent work \cite{2019arXiv190807320B}.  Explicit expressions of these matrix elements would have significant impact on the understanding of related quantities, for instance the Drude weight. Last but not least, the observed similarity between the two families of cumulants suggests that one could think of a  "free energy" that generates the time integrated currents in the same fashion that the usual TBA free energy generates  the conserved charges. 

The paper is structured as follows. In the first section we present basic ingredients of GHD and GGE. We also show how the first cumulants of conserved charges can be obtained from the GGE TBA free energy. In section 2 we first remind the technique of \cite{Kostov:2018ckg} to compute the partition function, or equivalently, to establish the TBA equation. We then use this technique to express the charge average and charge covariance as diagrams. Once the idea is clear,  we present the generalization to higher cumulants. In section 3.1 we remind the result of \cite{2018arXiv181202082M} for the total transport cumulants. We then show, up to the fourth cumulant that they can be equivalently represented by the same diagrams. Finally we comment on how the conjecture could be proven.

\section{GHD and GGE}
\subsection{Formulation of GHD}
In this section we present in more detail the quasi-particle formulation of GHD outlined in the introduction.

Consider an isolated, out of equilibrium integrable system.  After the relaxation time, a local steady state is reached at the mesoscopic-sized fluid cell centered around each point $x$ in space. This state is described by an infinite set of GGE chemical potentials $\Beta(x,t)=\lbrace\beta_1(x,t),\beta_2(x,t),...\rbrace$. The mean value of a local observable is given by
\begin{align}
\langle \mathcal{O}(x,t)\rangle =\frac{\tr [ e^{\sum_j -\beta_j(x,t) \Q_j}\mathcal{O}]}{\tr[ e^{\sum_j -\beta_j(x,t) \Q_j}]},\label{local-op-def}
\end{align}
where $\Q_j=\int dx Q_j(x,t)$ is an infinite set of conserved charges with density $Q_j(x,t)$, the integration runs over the volume of the cell.

The variation of GGE chemical potentials accross neighboring cells is small compared to the scale under consideration. In order to quantify this variation, one starts with the microscopic continuity equation
\begin{align}
\partial_t Q_j(x,t)+\partial_x J_j(x,t)=0.
\end{align}
Combined with the local equilibrium assumption, this leads to the
equation of state describing transport between neighboring cells
\begin{align}
\partial_t\langle Q(x,t)\rangle+ \partial_x\langle J(x,t)\rangle=0.
\end{align}
The average of the conserved charges can be obtained from GGE TBA as follows.

Let us consider for simplicity a theory with only one particle type. The particle energy and momentum are parametrized by the rapidity variable $\theta$: $E=m\cosh \theta$, $p=m\sinh\theta$. The conserved charges act diagonally on the basis of multi-particle wavefunctions 
\begin{align}
{\Q}_j|\theta_1,...,\theta_n\rangle =\sum_{i=1}^n q_j(\theta_i)|\theta_1,...,\theta_n\rangle\;.
\end{align} 
We restrict to "fermionic" case where these rapidities take different values. Particles undergo purely elastic scattering with two-to-two scattering phase $S(\theta,\eta)$ that depends on the difference of rapidities, we denote $K(\theta,\eta)=-i\partial_\theta \log S(\theta,\eta)$. The thermodynamic property of the state is encoded in the  particle density $\rho(x,t,\theta)$.  Space time dependence is implicitly understood from now. The Fermi-Dirac factor is parametrized by the so-called pseudo energy $\epsilon$: $f=\rho/(\rho+\rho_h)=1/(1+e^\epsilon)$, where $\rho_h$ denotes the density of holes. The pseudo-energy is the solution of the TBA equation
\begin{align}
\epsilon(\theta)=w(\theta)-\T\log[1+e^{-\epsilon}](\theta),\label{TBA-GGE}
\end{align}
where $w(\theta)=\sum_j\beta_jq_j(\theta)$ and $\T$ is the convolution with the scattering kernel normalized as follows
\begin{align}
\T\psi(\theta)\equiv \int\frac{d\eta}{2\pi}K(\theta,\eta)\psi(\eta)\;.\label{T-def}
\end{align}
TBA quantities are conveniently expressed in terms of the dressing operation
\begin{align}
\psi^\text{dr}=(1-\T f)^{-1}\psi\label{dressing}
\end{align}
which shows how a bare quantity gets renormalized by interaction.
In particular, the particle density is given by $\rho=f(p')^\text{dr}/(2\pi)$. Charge averages can either be written as the product of bare charge eigenvalue and particle density or vice versa
\begin{gather}
\langle Q_j(x,t)\rangle =\int d\theta \rho(\theta)q_j(\theta)=\int \frac{dp(\theta)}{2\pi} f(\theta) q_j^\text{dr}(\theta).\label{charge-density-average}
\end{gather}

The current average was conjectured in \cite{Castro-Alvaredo:2016cdj}
\begin{align}
\langle J_j(x,t)\rangle=\int d\theta v^\text{eff}(\theta)\rho(\theta)q_j(\theta),\quad \text{with}\quad v^\text{eff}\equiv \frac{(E')^\text{dr}}{(p')^\text{dr}}.\label{veff}
\end{align}
It was later proven in \cite{10.21468/SciPostPhys.6.2.023} for integrable quantum field theories and in \cite{2019arXiv190807320B} for spin chains.
The quantity $v^\text{eff}$ has the interpretation of effective velocity of quasi-particles propagating ballistically in local equilibrium.

The objects of study of this paper are the cumulants of conserved charges and those of  total currents. In the next section we remind how the former can be obtained from GGE free energy, we also discuss the advantages and drawbacks of this direct approach.

\subsection{Cumulants of conserved charges from GGE free energy}
By construction \eqref{local-op-def}, the cumulants of conserved charges are given by the derivatives of the free energy 
\begin{align}
F(\boldsymbol{\beta})=\log\tr[ e^{\sum_j -\beta_j(x,t) \Q_j}]= L\int\frac{dp(\theta)}{2\pi}\log[1+e^{-\epsilon(\theta)}]\label{GGE-free-energy}
\end{align}
with respect to the chemical potentials. Here $L$ denotes the volume of the fluid cell, which is large compared to the microscopic length scale of the theory. The charge average is simply
\begin{align}
\langle {\Q}_j\rangle=-\frac{\partial F}{\partial \beta_j}=L\int\frac{dp}{2\pi}f(\theta)\partial_{\beta_j} \epsilon(\theta).
\end{align}
The derivative of the pseudo-energy is nothing but the dressed charge eigenvalue $q_j^\text{dr}$. This can be seen by taking the derivative of the TBA equation \eqref{TBA-GGE} with respect to $\beta_j$. Therefore, the charge average can be written as
\begin{align}
\frac{1}{L}\langle {\Q}_j\rangle =\int \frac{dp}{2\pi}f(\theta)q_j^{\text{dr}}(\theta).\label{one-charge}
\end{align}
By translational invariance, this leads to the charge density average \eqref{charge-density-average}. The charge covariance matrix is given by 
\begin{align*}
\frac{1}{L}\langle {\Q}_j {\Q}_k\rangle^\text{c}&=\frac{\partial ^2 F}{\partial \beta_j\beta_k}=\int\frac{dp}{2\pi}\bigg\lbrace f(\theta)[1-f(\theta)]\partial_{\beta_k} \epsilon(\theta)\partial_{\beta_j} \epsilon(\theta)+ f(\theta)\partial_{\beta_k}\partial_{\beta_j} \epsilon(\theta)\bigg\rbrace.
\end{align*}
One can eliminate the second derivative of the pseudo energy 
\begin{align}
&\partial_{\beta_k}\partial_{\beta_j}\epsilon(\theta)=\int\frac{d\eta}{2\pi}K(\theta,\eta)\bigg\lbrace f(\eta)[1-f(\eta)]\partial_{\beta_k} \epsilon(\eta)\partial_{\beta_j} \epsilon(\eta)+f(\eta)\partial_{\beta_k}\partial_{\beta_j} \epsilon(\eta)\bigg\rbrace,\label{second-derivative}
\end{align}
by integrating this expression with the particle density measure. Using the fact that $\rho=f(p')^\text{dr}/(2\pi)$ one can then write the charge covariance  compactly as \cite{Mossel:2012vp,Doyon_2017}
\begin{align}
\frac{1}{L}\langle {\Q}_j {\Q}_k\rangle^\text{c}=\int du\rho(u)[1-f(u)]q_j^\text{dr}(\theta)q_k^\text{dr}(\theta).\label{GGE-2}
\end{align}
For the third cumulants, the same trick eliminates the third derivatives of the pseudo-energy but leaves the second ones
\begin{align}
\frac{1}{L} \langle {\Q}_j {\Q}_k {\Q}_l\rangle^\text{c}&=\int d\theta\rho(\theta)[1-f(\theta)][1-2f(\theta)]q_j^\text{dr}(\theta)q_k^\text{dr}(\theta)q_l^\text{dr}(\theta)\nonumber\\
&+\int d\theta\rho(\theta)[1-f(\theta)]\big[ q_j^\text{dr}\partial_{\beta_l}\partial_{\beta_k}\epsilon +q_j^\text{dr}\partial_{\beta_l}\partial_{\beta_k}\epsilon+q_j^\text{dr}\partial_{\beta_l}\partial_{\beta_k}\epsilon\big](\theta).\label{GGE-3}
\end{align}
This direct computation from free energy is clearly impractical for higher cumulants. There is no general rule to write the obtained expression in terms of fundamental TBA quantities like the particle density, the Fermi-Dirac factor or simple dressing operators. In the next section, we present a diagrammatic approach to compute cumulants of arbitrary order. The technique is based on the recent graph formulation of TBA \cite{Kostov:2018ckg,Kostov:2018dmi}.
\section{Diagrammatic formulation}
\subsection{Partition function as a sum over forests}
The first step in computing the partition function $
Z(\Beta)=\tr[ e^{\sum_j -\beta_j \Q_j}]$ is to choose a complete basis that diagonalizes all the conserved charges. We employ Bethe's hypothesis and index each multiparticle wavefunction by a set of quantum numbers
\begin{gather}
\phi_j=2\pi n_j,\quad \text{for integers }\; n_j,\quad j=\overline{1,N}.\label{Yang-Yang}
\end{gather}
The rapidity variables $\theta_j$ are related to the Bethe numbers $\phi_j$ through Bethe-Yang equations
\begin{gather}
\phi_j(\theta_1,...,\theta_N)=Lp(\theta_j)-i\sum_{k\neq j}^N\log S(\theta_j,\theta_k).\label{Bethe-Yang}
\end{gather}
At this point the partition function is written formally as a sum over mode numbers
\begin{align}
Z(\Beta)=\sum_{N\geq 0}\;\sum_{n_1<n_2<...<n_N}e^{-w(n_1,n_2,...,n_N)}.
\end{align}
In this expression, $w$ is an implicit function of mode numbers: one has to solve the Bethe-Yang equations \eqref{Bethe-Yang} for rapidities and replace
$ w(n_1,n_2,...,n_N)= w(\theta_1)+w(\theta_2)...+w(\theta_N)$. 

We would like to replace this discrete sum by an integral over phase space. First we have to remove the constraint among mode numbers \footnote{for "bosonic" and classical theories, see the end of this subsection}. This can be done by $1-\delta$ insertion. After expanding the Kronecker delta symbols, we obtain an unrestricted sum with multiplicities $(n_1^{r_1},...,n_N^{r_N})$. Such tuple defines an
(unphysical) Bethe state with $r_1+...+r_N$ particles. This state is a linear combination of plane waves with momenta $r_jp(\theta_j),\; j=1,...,N$ and thermal weight $
w(n_1^{r_1},...,n_N^{r_N})=r_1w(\theta_1)+...+r_Nw(\theta_N)$. 
The set of rapidities $\boldsymbol{\theta}$ is now given by Bethe-Yang equations with multiplicities
\begin{align}
\phi_j=Lp(\theta_j)-i\sum_{k\neq j}^N r_k\log S(\theta_j,\theta_k)+\pi(r_j-1)=2\pi n_j,\quad j=\overline{1,N}.\label{Yang-Yang-multiplicities}
\end{align}
The combinatorial factors due to Kronecker delta symbols being glued togther have been worked out in \cite{Kostov:2018dmi}
\begin{align}
Z(\Beta)=\sum_{N\geq 0}\frac{(-1)^N}{N!}\sum_{n_1,...,n_N\in\mathbb{Z}^N}\sum_{r_1,..,r_N\in\mathbb{N}^N}\prod_{j=1}^N\frac{(-1)^{r_j}}{r_j}e^{-w(n_1^{r_1},...,n_N^{r_N})}.
\end{align} 
Now we can transform this sum to an integral over rapidities through equations \eqref{Yang-Yang-multiplicities}
\begin{align*}
Z(\Beta)=\sum_{N\geq 0}\frac{(-1)^N}{N!}\sum_{r_1,..,r_n\in\mathbb{N}^N}\prod_{j=1}^N\frac{(-1)^{r_j}}{r_j}\int\frac{d\theta_1}{2\pi}...\frac{d\theta_N}{2\pi}\det G(\theta_1^{r_1},...,\theta_N^{r_N})e^{-r_1w(\theta_1)}....e^{-r_Nw(\theta_N)}.
\end{align*}
The Jacobian of this change of variables encodes all information about the interacting theory
\begin{align}
G_{kj}(\theta_1^{r_1},...,\theta_N^{r_N})=\big[Lp'(\theta_k)+\sum_{l\neq k} r_lK(\theta_k,\theta_l)\big]\delta_{kj}-r_kK(\theta_k,\theta_j)(1-\delta_{kj})\label{Gaudin}
\end{align}
Its determinant is also known as the Gaudin determinant that describes the norm of the state
\begin{align}
\det G(\theta_1^{r_1},...,\theta_N^{r_N})=\langle \theta_N^{r_N},...,\theta_1^{r_1}|\theta_1^{r_1},...,\theta_N^{r_N}\rangle.\label{Gaudin-norm}
\end{align}
In order to apply the matrix-tree theorem we consider the scaled matrix $\tilde{G}_{kj}=r_jG_{kj}$. This newly defined matrix is the sum of a diagonal matrix  and a Laplacian matrix 
\begin{gather*}
\tilde{G}_{kj}=\tilde{D}_k\delta_{jk}+\tilde{K}_{kj},\\
D_k=Lr_kp'(\theta_k),\quad \tilde{K}_{kj}=\delta_{kj}\sum_{l\neq k}r_kr_lK(\theta_k,\theta_l)-(1-\delta_{kj})r_kr_jK(\theta_k,\theta_j).
\end{gather*}
The matrix-tree theorem
then states that the determinant of $\tilde{G}$ is a sum over  forests $\mathcal{F}$ that span the totally
connected graph with vertices labeled by $j=1,...,N$. In each tree of the forest there is a vertex with the corresponding element of the matrix $D$ inserted. We define this vertex as the root of the tree
\begin{align}
\det \tilde{G}=\sum_{\mathcal{F}}\prod_{v_i \text{ roots}}D_i\prod_{\langle jk \rangle \;\text{branches}} r_jr_kK(\theta_j,\theta_k).
\end{align}
We can now write the partition function as
\begin{align*}
Z(\Beta)=\sum_{N\geq 0}\frac{(-1)^N}{N!}\sum_{r_1,..,r_n\in\mathbb{N}^N}\prod_{j=1}^N\int\frac{(-1)^{r_j}}{r_j^2}\frac{d\theta_j}{2\pi}e^{-r_jw(\theta_j)}\sum_{\mathcal{F}}\prod_{j \text{ roots}}Lr_jp'(\theta_j)\prod_{\langle jk \rangle } r_jr_kK(\theta_j,\theta_k).
\end{align*}

The next step is to invert the order of the sum over graphs and the integral/sum over the coordinates
$(\theta_j,r_j)$ assigned to the vertices. As a result we obtain a sum over the ensemble of
tree graphs, with their symmetry factors, embedded in the space $\mathbb{R}\times \mathbb{N}$ where the coordinates $(\theta,r)$
of the vertices take values. The embedding is free, in the sense that the sum over the positions of
the vertices is taken without restriction. As a result, the sum over the embedded tree graphs is the
exponential of the sum over connected ones. One can think of these graphs as tree level Feynman
diagrams obtained by applying the following Feynman rules

\begin{equation}
 \boxed{\begin{split} \label{Feynmp} 
 \\
 \imineq{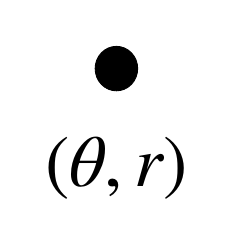}{8} \  \qquad  &= \  \  \ {(-1)^{r-1} \over r^2} \ e^{-r w(\theta)}
\\
 \imineq{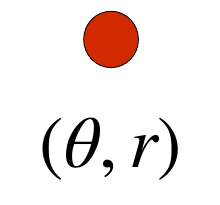}{7} \ \ \qquad  &= \  \  \  Lr p'(\theta)\, {(-1)^{r-1} \over r^2} \ e^{-r w(\theta)}
\\
 \imineq{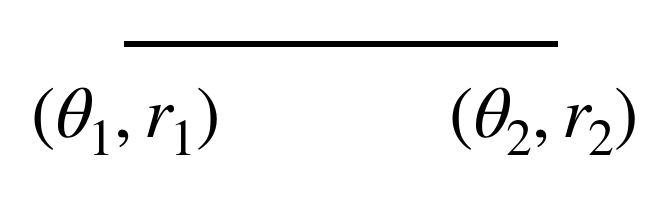}{6.5} \hskip-3mm  \  \ \ &= \ \ \ r_1 r_2 K(\theta_2, \theta_1)
 \end{split} 
 }
\end{equation}
In this way we can write the free energy as
\begin{align}
F(\beta)=\int \frac{dp(\theta)}{2\pi}\sum_{r\geq 1}rY_r(\theta),
\end{align}
where $Y_r(\theta)$ is the sum of all trees rooted at the point $(\theta,r)$. As any partition sum of trees, it satisfies a simple non-linear equation (a Schwinger-Dyson equation in the QFT language)
\begin{align}
Y_r(\theta)=\frac{(-1)^r}{r^2}\big[e^{-w(\theta)}\exp\sum_{s} \int\frac{d\eta}{2\pi}sK(\eta,\theta)Y_s(\eta)\big]^r\label{Schwinger-Dyson}
\end{align}
which translates the following combinatorial structure of trees
\begin{figure}[h]
\centering
\includegraphics[width=15cm]{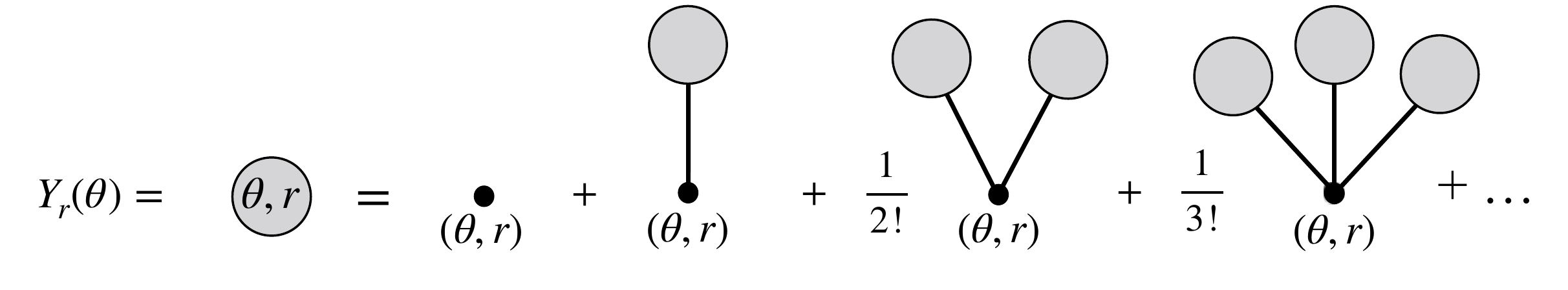}
\caption{Diagrammatic representation of the TBA equation}
\end{figure}

In particular  $Y_r(\theta)=(-1)^rY_1^r(\theta)/r^2$, and the equation \eqref{Schwinger-Dyson} for $r=1$ is the usual TBA equation  with $Y_1(\theta)=Y(\theta)=e^{-\epsilon(\theta)}$
\begin{gather*}
Y(\theta)=e^{-w(\theta)}\exp\int\frac{d\eta}{2\pi}K(\eta,\theta)\log[1+Y(\eta)],\quad F(\Beta)=L\int\frac{dp(\theta)}{2\pi}\log[1+Y(\theta)].
\end{gather*}
Our machinery is robust: by fixing all the signs to positive, we obtain the bosonic TBA; by discarding the multiplicities we obtain the TBA of classical particles. The convoluted terms in the TBA equation of these theories are respectively $-\log(1-Y)$ and $Y$.

In the following sections we address the cumulants of conserved charges. The main idea is to combine the normalization of states \eqref{Gaudin-norm} with the fact that all conserved charges act diagonally on these states. We begin with the charge average.
\subsection{Charge average}
As the conserved charges act diagonally on the multi-particle wavefunctions
\begin{align}
\langle \theta_N^{r_N},...,\theta_1^{r_1}|{\Q}_j|\theta_1^{r_1},...,\theta_N^{r_N}\rangle=\sum_{i=1}^N r_iq_j(\theta_i)\langle \theta_N^{r_N},...,\theta_1^{r_1}|\theta_1^{r_1},...,\theta_N^{r_N}\rangle, \label{charge-multistate}
\end{align}
we can evaluate the nominator of the expression
\begin{align*}
\langle {\Q}_j\rangle _{\Beta}=\frac{\Tr [e^{-\sum \beta _i \Q_i}\Q_j]}{\Tr [e^{-\sum \beta_i{\Q}_i}]}
\end{align*}
following the above steps. The result in a sum over forests with the same Feynman rules as  \eqref{Feynmp}. The only modification is that one vertex $(\theta,r)$ of the forest is marked with a charge insertion and carries an extra weight of $rq_j(\theta)$ coming from \eqref{charge-multistate}. Contribution coming from un-inserted trees (vacuum diagrams) cancel with the denominator. As a result, the average of $\Q_j$ is a sum over trees with a vertex marked with charge insertion. From this vertex one can always trace a unique path to the root of the tree.
\begin{figure}[h]
\centering
\includegraphics[width=15cm]{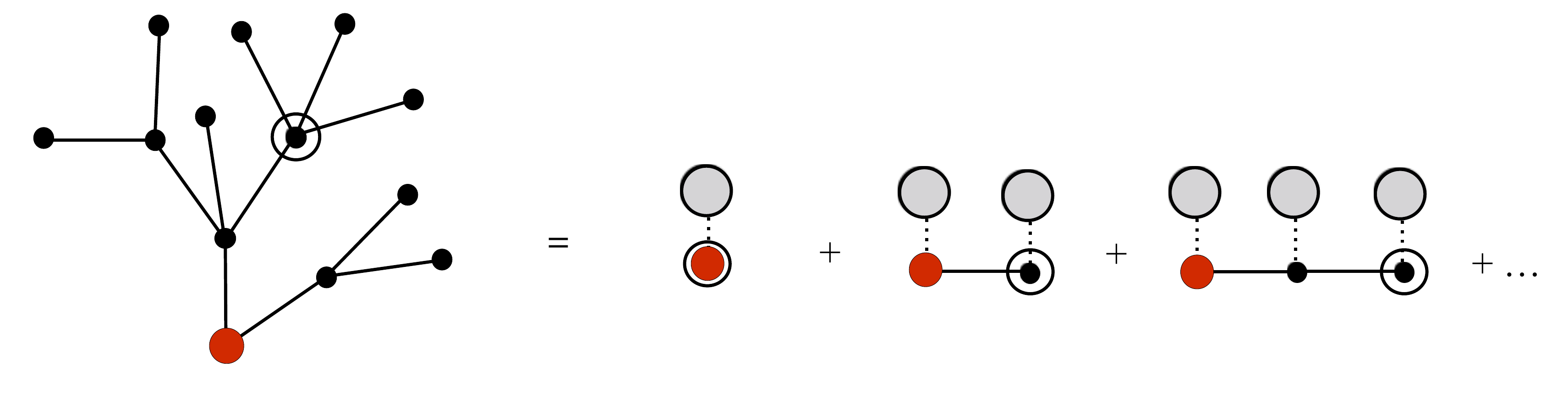}
\caption{Diagrammatic representation of the charge average}
\end{figure}

At each node $(\theta,r)$ inside this spine, one can sum up the trees growing out of it while pulling the multiplicities $r^2$ from the two adjacent propagators. The nodes at the two ends of the spine receive a multiplicity from the charge (or momentum derivative) insertion and a residual multiplicity from one propagator. This results in the Fermi-Dirac factor on every nodes along the spine
\begin{align*}
\sum_{r\geq 1} r^2Y_r(\theta)=\sum_{r\geq 1}(-1)^{r-1}Y^r(\theta)=\frac{Y(\theta)}{1+Y(\theta)}=f(\theta).
\end{align*}
Moreover, the sum over spines each of which carries a Fermi-Dirac factor on its nodes is nothing but the dressing operation \eqref{dressing}. We recover the expression  \eqref{one-charge} of the charge average
\begin{align}
\frac{1}{L}\langle {\Q}_j\rangle=\int \frac{dp}{2\pi}f(\theta)q_j^\text{dr}(\theta).\label{LM-conserved}
\end{align}

\subsection{Charge covariance}
Consider now two conserved charges $\Q_j$ and $\Q_k$. In evaluating the average of their product
\begin{align*}
\langle {\Q}_j{\Q}_k\rangle =\frac{\Tr [e^{-\sum \beta _i \Q_i}{\Q}_j{\Q}_k]}{\Tr [e^{-\sum \beta _i \Q_i}]}
\end{align*}
the insertion of a complete basis is factorized
\begin{align}
\langle \theta_N^{r_N},...,\theta_1^{r_1}|{\Q}_j{\Q}_k|\theta_1^{r_1},...,\theta_N^{r_N}\rangle =\langle \theta_N^{r_N},...,\theta_1^{r_1}|\theta_1^{r_1},...,\theta_N^{r_N}\rangle\sum_{i=1}^N r_iq_j(\theta_i)\sum_{i=1}^N r_iq_k(\theta_i).\label{2charge-multistate}
\end{align}
As a consequent, we obtain a sum over forests in which the two charges are inserted at two arbitrary vertices. Un-inserted sub-forests again cancel with the denominator. The two charges can fall on the same tree or on two different trees. The sum over the latter is factorized into the two charge averages.  Therefore the charge covariance is given by the sum over trees with two charge insertions
\begin{figure}[h]
\centering
\includegraphics[width=6cm]{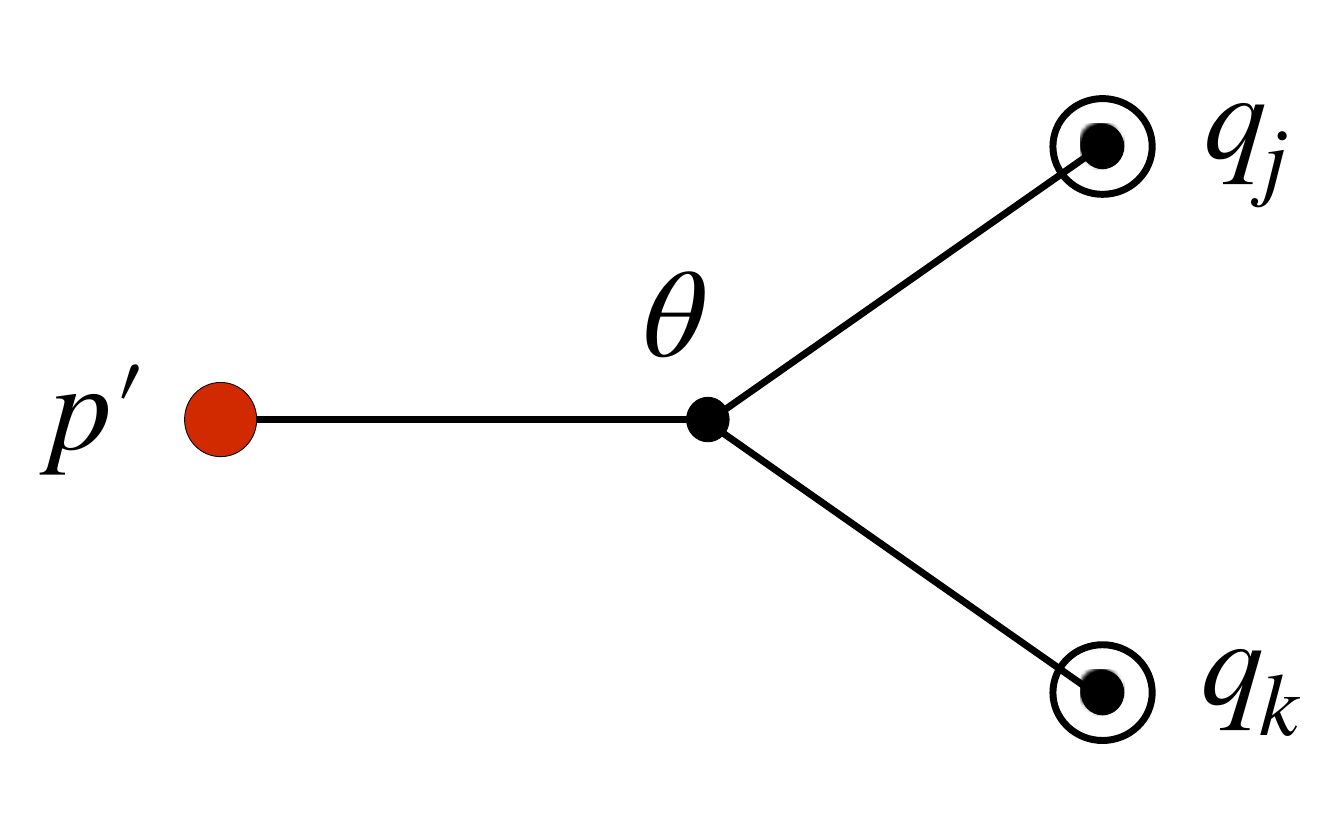}
\caption{Combinatorial structure of a tree with two leaves: there exists an internal vertex connected to the three external ones. We take this vertex as a reference point to sum over the trees.}\label{2-charges}
\end{figure}

From each charge-inserted vertex, one can find a unique path down to the root of the tree. The two paths must join at some point $(\theta,r)$: a unique vertex linked to the root and the two leaves. Except for this vertex, all other vertices receive the Fermi-Dirac factor $f$ as explained above. At this vertex we can pull three multiplicities from the three adjacent propagators. This results in a special filling factor
\begin{align*}
\sum_{r\geq 1} r^3Y_r(\theta)=\frac{Y(\theta)}{[1+Y(\theta)]^2}=f(\theta)[1-f(\theta)].
\end{align*}
The charge covariance involves three dressed quantities corresponding to the three spines coming out of this intersection point. We recover the expression \eqref{GGE-2}
\begin{align}
\frac{1}{L}\langle {\Q}_j{\Q}_k \rangle^\text{c}=\int\frac{d\theta}{2\pi}f(\theta)[1-f(\theta)](p')^\text{dr}(\theta)q_j^\text{dr}(\theta)q_k^\text{dr}(\theta).\label{two-point}
\end{align}

\subsection{Higher cumulants}

After understanding the explicit examples of the charge average and charge covariance, generalization to higher cumulants is straightforward. The $n^\text{th}$ cumulant is given by a sum over all tree-diagrams with $n+1$ external vertices : a root with $p'$ inserted and $n$ leaves carrying the $n$ conserved charges. Their internal vertices live in  phase space and will be integrated over. An external propagator connecting an internal vertex $\theta$ and an external vertex with an operator $\psi$ is assigned a weight $\psi^\text{dr}(\theta)$, here $\psi$ can either be the momentum derivative or the charges. An internal propagator connecting two internal vertices $\theta,\eta$ has a weight $K^\text{dr}(\theta,\eta)$, where 
\begin{align}
K^\text{dr}(u,v)=K(u,v)+\int\frac{dw}{2\pi}K(u,w)f(w)K^\text{dr}(w,v).\label{dressed-propagator}
\end{align}  
An internal vertex $\theta$ of degrees $d$ has a weight
\begin{align*}
\sum_{r\geq 1}(-1)^{r-1}r^{d-1}Y^r(\theta).
\end{align*}
We summarize these rules in the following
\medskip \medskip 
\begin{equation}
 \boxed{\begin{split} \label{reFeynman} 
 \\
 \imineq{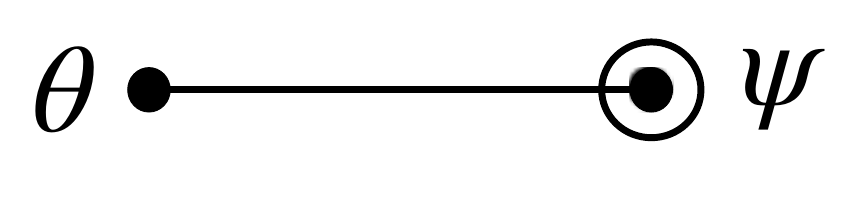}{4} \  \qquad  &= \  \  \ \psi^{\text{dr}}(\theta)
\\
 \imineq{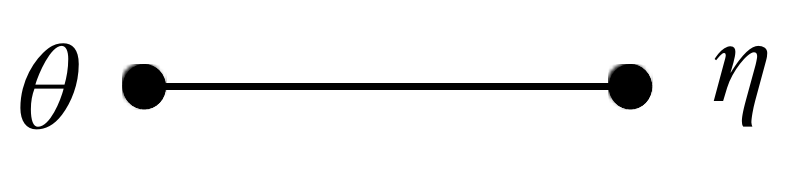}{4}  \ \qquad  &= \  \  \  K^\text{dr}(\theta,\eta)
\\
 \imineq{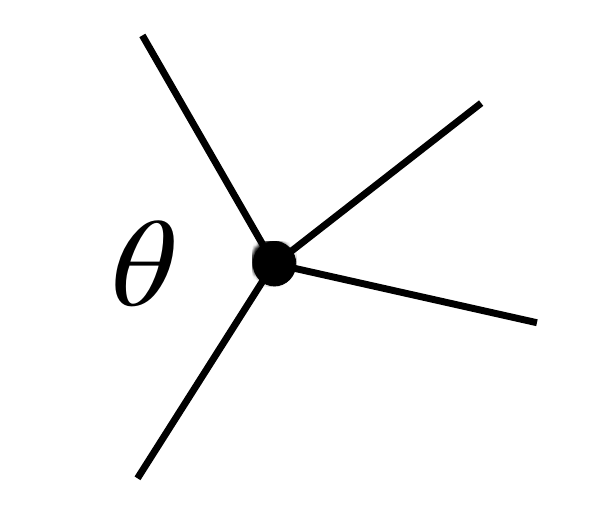}{11} \hspace{1cm}  \  \ \ &= \ \ \ \sum_{r\geq 1}(-1)^{r-1}r^{d-1}Y^r(\theta)
 \end{split} 
 }
\end{equation}
There is a simple recursive algorithm to generate all diagrams with $n$ leaves. For each partition of $n$ that is not the trivial one ($n=n$)
\begin{align}
n=\underbrace{a_1+...+a_1}_{\alpha_1}+\underbrace{a_2+...+a_2}_{\alpha_2}+...+\underbrace{a_j+...+a_j}_{\alpha_j},\quad a_1<a_2<...<a_j
\end{align} 
we choose $\alpha_1$ trees with $a_1$ leaves, ..., $\alpha_j$ trees with $a_j$ leaves. We then remove their roots and join them to a new common root. This algorithm translates into the following equation that determines the number $d_n$ of diagrams with $n$ leaves
\begin{align}
d_n=\sum_{\substack{p\in \mathcal{P}_n,\; |p|>1\\
p=(a_1^{\alpha_1},...,a_j^{\alpha_j})}}\prod_{i=1}^j\binom{d_{a_1}+\alpha_i-1}{\alpha_i}.
\end{align}
Some values of $d_n$ are given in the following table
\begin{table}[!htb]
\centering
  \begin{tabular}{ |c|c | c | c|c|c|c|c|c|c|c| }
    \hline
    $n$& $1$ & $2$ & $3$& $4$ & $5$ & 6 & 7 &8&9&10 \\ \hline
    $ d_n$& $1$ & $1$ & $2$& $5$ & $12$ & 33& 90& 261 & 766 & 2312\\ \hline
  \end{tabular}
\end{table}

\begin{figure}
\centering
\includegraphics[width=13cm]{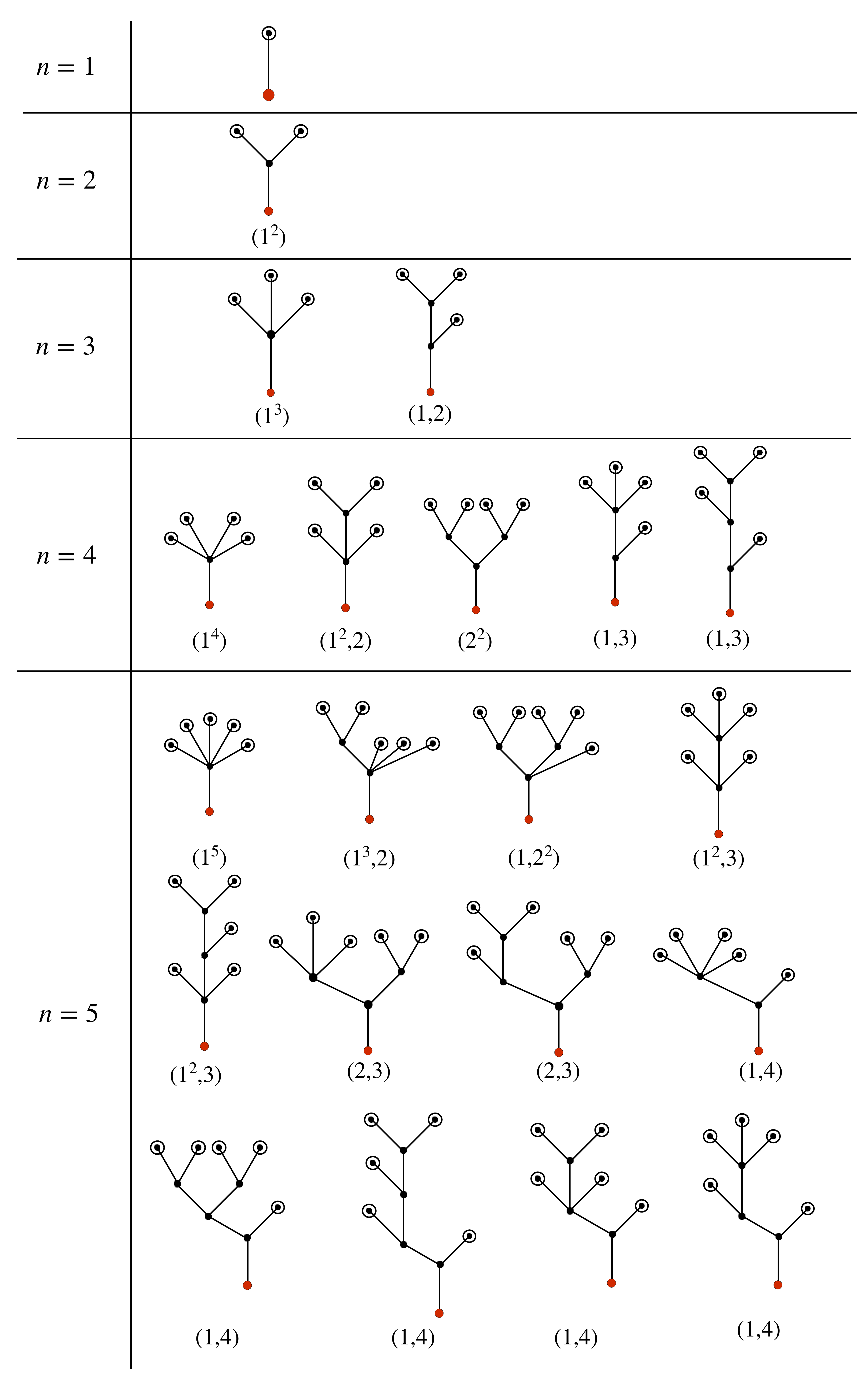}
\caption{Trees up to five leaves along with the partition used to generate them.}
\label{many}
\end{figure}

We also list all diagrams with up to $5$ leaves in figure \ref{many}.

The third cumulant $\langle  {\Q}_j{\Q}_k{\Q}_l\rangle ^\text{c}$
can be read directly from the two diagrams with three leaves. The one on the left gives
\begin{align}
\int\frac{d\theta}{2\pi} f(\theta)[1-f(\theta)][1-2f(\theta)](p')^{\text{dr}}(\theta)q_j^{\text{dr}}(\theta)q_k^{\text{dr}}(\theta)q_l^{\text{dr}}(\theta).\label{second-diagram}
\end{align}
The one on the right involves three permutations of vertices
\begin{align}
\int\int\frac{d\theta}{2\pi}\frac{d\eta}{2\pi} (p')^{\text{dr}}(\theta)f(\theta)[1-f(\theta)] f(\eta)[1-f(\eta)]K^{\text{dr}}(\eta,\theta)\big[q_j^{\text{dr}}(\theta)q_k^{\text{dr}}(\eta)q_l^{\text{dr}}(\eta)\nonumber\\
+q_k^{\text{dr}}(\theta)q_j^{\text{dr}}(\eta)q_l^{\text{dr}}(\eta)+q_l^{\text{dr}}(\theta)q_j^{\text{dr}}(\eta)q_k^{\text{dr}}(\eta)\big].\label{three-first}
\end{align}
This result agrees with expression \eqref{GGE-3} obtained from GGE free energy. Indeed, we can write the second derivative of pseudo-energy \eqref{second-derivative} in terms of the dressed propagator as follows
\begin{align*}
\partial_{\beta_l}\partial_{\beta_k}\epsilon(\theta)=\int\frac{d\eta}{2\pi} f(\eta)[1-f(\eta)] K^{\text{dr}}(\theta,\eta)\partial_{\beta_k}\epsilon(\eta)\partial_{\beta_l}\epsilon(\eta).
\end{align*}
\section{Cumulants of the total transport in GHD}
In this section we restrict our discussion to stationary states. A prototypical example is the one arising at the junction of the partitioning protocol: two systems (left and right) thermalized at different temperatures being brought to contact and let evolved unitarily. The occupation number of this state can be deduced \cite{Castro-Alvaredo:2016cdj} from the initial condition 
\begin{align}
f(\theta)=f^{\text{L}}(\theta)\Theta(\theta-\theta^*)+f^{\text{R}}(\theta)\Theta(\theta^*-\theta).\label{bipartie}
\end{align} 
In this equation, $f^{\text{L,R}}$ are respectively the occupation numbers of the asymptotic left and right subsystem and $\Theta$ is the Heavyside function. The transition point $\theta^*$ is implicitly determined by requiring that the ballistic propagation velocity there vanishes $v^\text{eff}(\theta^*)=0$. To remind, this effective velocity is obtained from the occupation profile $f$ via \eqref{veff}.

We then consider  the cumulants of the total flow crossing this point scaled with inverse time
\begin{align}
\lim_{t\to\infty}\frac{1}{t}\int_0^t dt_1...\int_0^t dt_n \langle J_1(0,t_1)...J_n(0,t_n)\rangle ^\textnormal{c}.\label{J-n}
\end{align}
These cumulants play a central role in the large deviation theory as they characterize the probabilities of rare events with significant deflection from their mean values \cite{TOUCHETTE20091}.
We conjecture that they are given by the same diagrams presented in the previous section, with only two modifications: the operator  at the root is the energy derivative $E'$ (instead of the momentum derivative) and each internal vertex $\theta$ of odd degree carries an extra sign of the effective velocity $\sgn[v^\text{eff}(\theta)]$. 

We confirm our conjecture by a non-trivial matching with the result of \cite{Doyon:2019osx} up to the fourth cumulant of the same current. We first remind how this quantity was derived in \cite{SciPostPhys.3.6.039} and\cite{2018arXiv181202082M,Doyon:2019osx}.
\subsection{Hydrodynamic approximation}
The covariance matrix was first studied in \cite{SciPostPhys.3.6.039} and named the "Drude self-weight" 
\begin{align*}
D^\textnormal{s}_{ij} \equiv \lim_{t\to \infty}\int_0^t ds\langle J_i(0,s)J_j(0,0)\rangle^\textnormal{c}.
\end{align*}
The name comes from its resemblance with the conventional Drude weight
\begin{align*}
D_{ij}\equiv \lim_{t\to \infty}\int dx \langle J_i(x,t)J_j(0,0)\rangle^\textnormal{c}.
\end{align*}
The conventional Drude weight characterizes the zero-frequency conductancy in integrable systems and controls important transport properties \cite{PhysRevLett.119.020602}. It was shown in \cite{SciPostPhys.3.6.039} that the two quantities actually share similar expressions. If one can deduce from our conjecture the matrix element of the Drude self-weight, the same quantity for the Drude weight could potentially be obtained.

The derivation of \cite{SciPostPhys.3.6.039} relies on the current sum rule \cite{Mendl_2015} that allows the Drude self-weight to be expressed in terms of the charge-charge correlation function
\begin{align}
D^\textnormal{s}_{ij}=\int dx |x|\frac{1}{2}\big[\langle Q_i(x,t)Q_j(0,0)\rangle^\text{c}+\langle Q_j(x,t)Q_i(0,0)\rangle^\text{c}\big]
\end{align}
and the large distance limit of this correlator \cite{SciPostPhys.1.2.015}. The result can be written as
\begin{align}
D^\textnormal{s}_{ij}=\int \frac{d\theta}{2\pi} (E')^\text{dr}(\theta)s(\theta)f(\theta)[1-f(\theta)]q_i^\text{dr}(\theta)q_j^\text{dr}(\theta),\label{Drude-s}
\end{align}
where we have denoted for short $s(\theta)=\sgn[v^\text{eff}(\theta)]$.

In \cite{2018arXiv181202082M,Doyon:2019osx} all the diagonal cumulants were studied at once by mean of their generating function
\begin{align}
F(\lambda)=\sum_{n=1}^\infty\frac{\lambda^n}{n!}c_n\quad \textnormal{with}\quad c_n=\lim_{t\to\infty}\frac{1}{t}\int_0^t dt_1...\int_0^t dt_n \langle J(0,t_1)...J(0,t_n)\rangle ^\textnormal{c}\label{gen-function}
\end{align}
A functional equation satisfied by this function has been found by fluctuations from Euler-scale hydrodynamics. From this equation one can derive an explicit expression for each cumulant $c_n$ for any value of $n$. Nevertheless, such derivation requires special manipulation for each case. It seems  possible however that the individual cumulants can be derived from the same principle without considering the generating function, see the discussion at the end of this section. In the following we remind the result of \cite{Doyon:2019osx} for $c_{2,3,4}$ and show that they possess the same combinatorial structure as the  cumulants of the corresponding conserved charges.

The authors of \cite{Doyon:2019osx} also considered a generating function with different variables. Establishing a functional equation for such function would lead to non-diagonal cumulants. It would be interesting to see if this approach is in agreement with our conjecture.
\subsection{Comparison with diagrams} 
In this section we show that the result of \cite{2018arXiv181202082M} is correctly reproduced by our diagrams.

The Drude self-weight is given by \eqref{Drude-s} and can be represented as the diagram in figure \ref{2-charges} with energy derivative at its root and the sign of the effecive velocity at its internal vertex (of degree 3).

The third cumulant was found to be
\begin{align}
c_3=\int\frac{d\theta}{2\pi} (E')^\text{dr}(\theta)f(\theta)&[1-f(\theta)]s(\theta)q^\text{dr}(\theta)\times \nonumber \\
&\times \big\lbrace [1-2f(\theta)][q^\text{dr}(\theta)]^2s(\theta)+3\big[(q^\text{dr})^2(1-f)s\big]^{*\text{dr}}(\theta)\big\rbrace,\label{third-cumulant}
\end{align}
where the star-dressing operator is defined as 
\begin{gather}
\psi^{*\text{dr}}(\theta)\equiv \psi^\text{dr}(\theta)- \psi(\theta) .\label{star-dr-def}
\end{gather}
\begin{figure}[h]
\centering
\includegraphics[width=13cm]{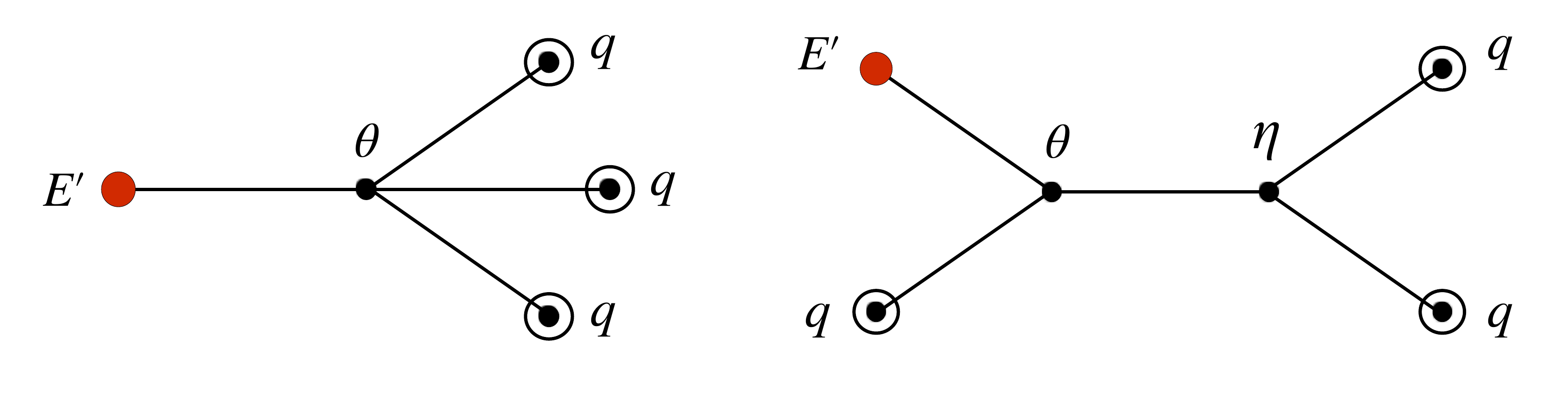}
\caption{Two trees with three leaves}
\label{three-leaves}
\end{figure}
The first term of \eqref{third-cumulant} is given by the left diagram in figure \ref{three-leaves}, again with energy derivative at the root. The internal vertex of this diagram is of degree 4 so there is no sign of the effective velocity. The second term is given by diagram on the right which comes with a symmetry factor of 3. The internal vertices are both of degrees 3 so each comes with a sign of the effective velocity. The matching is easily seen with the following writing of the star dressing operator in terms of the dressed propagator \eqref{dressed-propagator}
\begin{gather}
\psi^{*\text{dr}}(\theta)=\int\frac{d\eta}{2\pi}K^{\text{dr}}(\eta,\theta)f(\eta)\psi(\eta).\label{stardressing-equiv}
\end{gather}
This identity also reveals the physical picture behind our diagrams: the integration over internal vertices is nothing but the contribution from virtual particles that carry anomalous corrections  to the bare charges.

The fourth cumulant is considerably more complicated and constitutes a highly non-trivial check for our conjecture. The original formula of  $c_4$ as it was derived in \cite{2018arXiv181202082M} is 
\begin{align}
&c_4=\int\frac{d\theta}{2\pi}(E')^\text{dr}(\theta)f(\theta)[1-f(\theta)]\times \bigg\lbrace \frac{Y(\theta)^2+6Y(\theta)+6}{[Y(\theta)+1]^2}s(\theta)[q^\text{dr}(\theta)]^4\nonumber\\
+&3s(\theta)\lbrace [(1-f)s(q^\text{dr})^2]^\text{dr}(\theta)\rbrace^2+12s(\theta)q^\text{dr}(\theta)\lbrace(1-f)sq^\text{dr}[(1-f)s(q^\text{dr})^2]^{\text{dr}}\rbrace^{\text{dr}}(\theta)\nonumber\\
+&6[f(\theta)-2)[q^\text{dr}(\theta)]^2[s(1-f)(q^\text{dr})^2]^\text{dr}(\theta)+4s(\theta)q^\text{dr}(\theta)[(1-f)(f-2)(q^\text{dr})^3]^\text{dr}(\theta)\bigg\rbrace.\label{c4}
\end{align}
For convenience, we repeat here all diagrams with four leaves 
\begin{figure}[h]
\centering
\includegraphics[width=12cm]{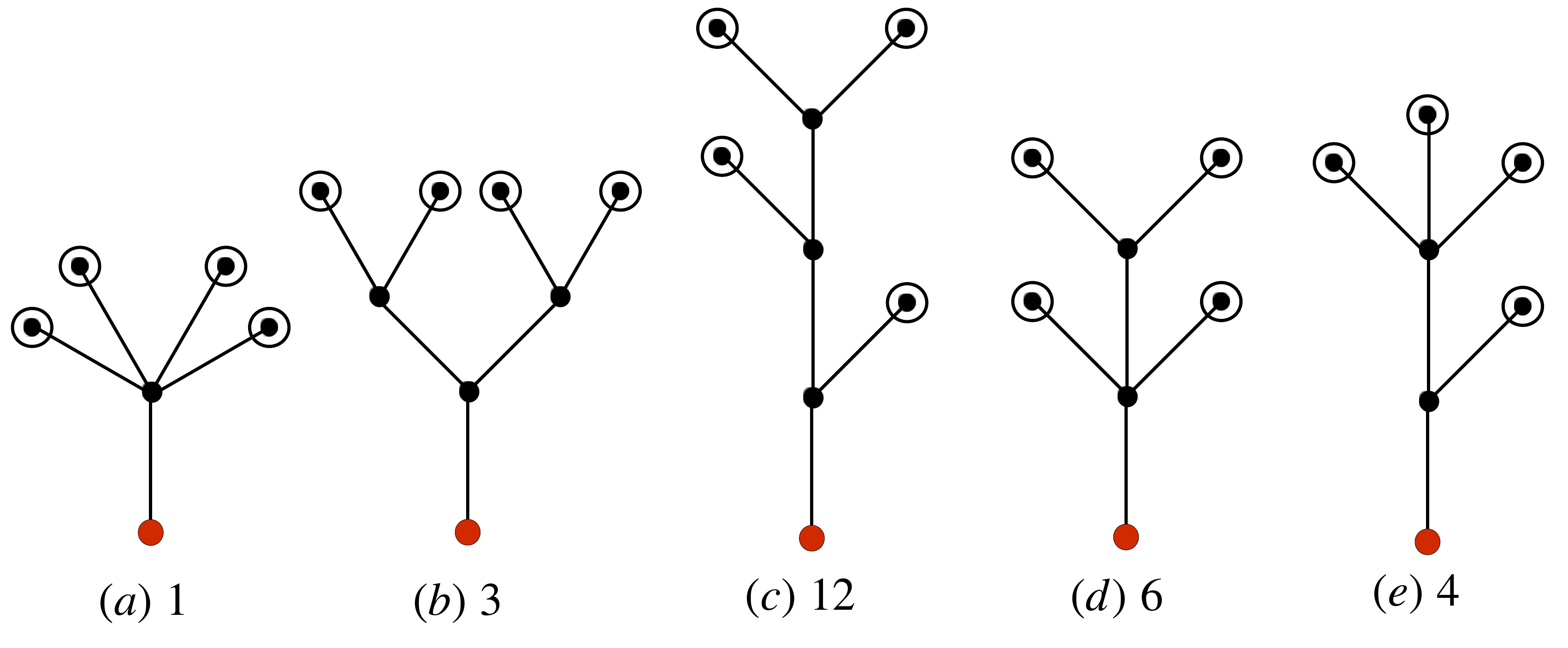}
\caption{Trees with four leaves along with their symmetry factors}
\label{4-tree}
\end{figure}

Due to the identity \eqref{stardressing-equiv}, our trees are naturally expressed in terms of the star dressing operation. In order to compare them with \eqref{c4}, we repeatedly use the definition \eqref{star-dr-def} to make appear the dressing operation. We then show that the discrepancies cancel each other. The integration variable $\theta$ in the formula \eqref{c4} corresponds to the coordinate of the internal vertex closest to the root of each tree. These vertices are always of degree at least 3, therefore we can factorize a factor $f(\theta)[1-f(\theta)]$ from their weights. After this factorization, the contribution from the trees are (we omit the dependence on $\theta$)
\begin{itemize}
\item Tree (a)
\begin{align*}
\frac{Y^2-4Y+1}{(Y+1)^2}s(q^\text{dr})^4=\frac{Y^2+6Y+6}{(Y+1)^2}s(q^\text{dr})^4-5(1-f^2)s(q^\text{dr})^4
\end{align*}
\item Tree (b)
\begin{align*}
&3s\lbrace[(1-f)s(o^\text{dr})^2]^{*\text{dr}}\rbrace^2=3s \lbrace [(1-f)s(o^{\text{dr}})^2]^{\text{dr}}\rbrace^{2}\\
+&3s(1-f)^2(o^\text{dr})^4-6(1-f)(o^\text{dr})^2[(1-f)s(o^\text{dr})^2]^\text{dr}
\end{align*}

\item Tree (c)
\begin{align*}
&12sq^\text{dr}\lbrace(1-f)sq^\text{dr}[(1-f)s(q^\text{dr})^2]^{*\text{dr}}\rbrace^{*\text{dr}}=12sq^\text{dr}\lbrace(1-f)sq^\text{dr}[(1-f)s(q^\text{dr})^2]^{\text{dr}}\rbrace^{\text{dr}}\\
-&12(1-f)(q^\text{dr})^2[(1-f)s(q^\text{dr})^2]^\text{dr}-12sq^\text{dr}[(1-f)^2(q^\text{dr})^3]^\text{dr}+12(1-f)^2s(q^\text{dr})^4
\end{align*} 
\item Tree (d)
\begin{align*}
&6(1-2f)(q^\text{dr})^2[(1-f)s(q^\text{dr})^2]^{*\text{dr}}=6(f-2)(q^\text{dr})^2[(1-f)s(q^\text{dr})^2]^{\text{dr}}\\
-&6s(f-2)(1-f)(q^\text{dr})^4+18(1-f)(q^\text{dr})^2[(1-f)s(q^\text{dr})^2]^{\text{dr}}-18s(1-f)^2(q^\text{dr})^4
\end{align*}
\item Tree (e)
\begin{align*}
&4sq^\text{dr}[(1-f)(1-2f)(q^\text{dr})^3]^{*\text{dr}}=4sq^\text{dr}[(1-f)(f-2)(q^\text{dr})^3]^\text{dr}\\
-&4s(q^\text{dr})^4(1-f)(f-2)-12s(q^\text{dr})^4(1-f)^2+12sq^\text{dr}[(1-f)^2(q^\text{dr})^3]^\text{dr}
\end{align*}
\end{itemize}
The discrepancies indeed cancel each other.
\subsection{Comments on the conjecture}
There are two plausible ways to prove our conjecture.

First, one can try to derive the matrix elements of the product of total currents. One can then repeat the same steps of section 2 to perform their summation. The correct matrix elements must guarantee that the resulting diagrams have energy derivative at their root and sign of the effective velocity at their odd internal vertices. Concerning these two properties, the former is expected while the latter is more puzzling. Let us elaborate on this point.

In our proof of the current average \cite{10.21468/SciPostPhys.6.2.023} it was understood that the form factor of a current is very similar to that of the corresponding charge: both are given by trees, the only difference being the operator at the root. It is then natural that any average involving currents, if admits combinatorial structure of trees, would have the energy derivative at the roots.

As for the sign of the effective velocity, a naive guess would be to assign such sign for each bare propagator and for each external vertex. Most of them will cancel each other except for internal vertices of odd degrees. The flaw in this argument is that the weights of graph components should involve only bare quantities, like the ones in \eqref{Feynmp}. Only after the graphs are summed over do we have  renormalized (dressed) quantities, see \eqref{reFeynman}. The effective velocity is a dressed quantity and as such cannot be included in the  weight of bare propagators. In most cases however, the sign of the effective velocity coincides with that of the rapidity and the above modification could in principle be implemented.

Second, one can regard the combinatorial structure of the charge cumulants as a result of successive derivatives on the free energy \eqref{GGE-free-energy}. Simply speaking these derivatives generate branches and joints (internal vertices) of the trees. If one can prove the existence of a similar "free  energy" whose derivatives lead  to cumulants of the total transport, it is natural that the same combinatorial structure would arise. Such free energy should not be confused with the generating function \eqref{gen-function}: what we seek for is the derivative with respect to the GGE chemical potentials, not the auxiliary variable $\lambda$.

This approach seems possible in view of the following identity, proven in \cite{Doyon:2019osx}
\begin{align}
\int_0^t ds\langle J_i(0,s) \mathcal{O}(0,0)\rangle ^\text{c}=-\sum_{j}\sgn(A)_{ij}\frac{\partial}{\partial\beta_j}\langle \mathcal{O}(0,0)\rangle
\end{align}
for any local observable $\mathcal{O}$. Here $A$ is the flux Jacobian matrix $
A_{ij}=\partial \langle J_i(0,0)\rangle/\partial \langle Q_j(0,0)\rangle$, 
and the sign is defined as the sign matrix of its eigenvalues. If one can show that this identity is still valid when the local operator $\mathcal{O}$ is replaced by the product of the total currents then one would be able to obtain their cumulants from successive derivatives of the current average. 

\section*{Conclusion}
In this paper we present a new approach that allows the cumulants of conserved charges in a GGE to be written as a sum over simple diagrams. The weights of these diagrams are readily obtained from TBA data. Our formalism provides an intuitive picture of these cumulants: external vertices are the bare charges while internal vertices are virtual particles that carry anomalous corrections. We also conjecture that the same diagrams, with minor modifications, describe the cumulants of total transport currents in GHD. We confirm our conjecture by a non-trivial matching with the result of \cite{2018arXiv181202082M} obtained by linear fluctuating dynamics.

In future work, we would like to see the extend of this combinatorial structure in dynamical correlation functions and related quantities. The study of large scale correlation functions in GHD has been addressed in \cite{Doyon:2017vfc}. For the charge-charge and charge-current correlation functions, the same combinatorial structure continues to hold, with the inclusion of a space-time propagator.  The situation is more subtle for the current-current correlator and the Drude weight. These quantities involve  the inverse of a dressed quantity  and it is currently not clear how such inversion could be represented in our formalism.
\section*{Acknowledgment}
The author thanks Takato Yoshimura for introduction to this problem, for early collaboration and important comments on the draft. The author also thanks Benjamin Doyon and Jason Myers for helpful discussions and clarification of their results. The last part of this work was done under the hospitality of Osaka University and the generous support of Asia Pacific Center for Theoretical Physics.

The author is supported by Le Minist\`ere de l’Enseignement Sup\'erieur et de la Recherche et de l’Innovation.

%\bibliography{two-point}
%\bibliographystyle{unsrt}
\end{document}